\pdfoutput=1
\documentclass[11pt]{article}

\usepackage[]{acl}

\usepackage{times}
\usepackage{latexsym}

\usepackage[T1]{fontenc}
\usepackage[utf8]{inputenc}

\usepackage{microtype}
\usepackage{graphicx}
\usepackage{enumitem}
\usepackage{caption}
\usepackage{subcaption}
\usepackage{amsfonts}
\title{How Much Does Prosody Help Turn-taking? \linebreak Investigations using Voice Activity Projection Models} 
 
\author{Erik Ekstedt \\
  KTH Speech, Music and Hearing \\
  Stockholm, Sweden \\
  \texttt{erikekst@kth.se} \\\And
  Gabriel Skantze\\
  KTH Speech, Music and Hearing \\
  Stockholm, Sweden \\
  \texttt{skantze@kth.se} \\}

\begin{document}
\maketitle

\begin{abstract}
Turn-taking is a fundamental aspect of human communication and can be described as the ability to take turns, project upcoming turn shifts, and supply backchannels at appropriate locations throughout a conversation. In this work, we investigate the role of prosody in turn-taking using the recently proposed Voice Activity Projection model, which incrementally models the upcoming speech activity of the interlocutors in a self-supervised manner, without relying on explicit annotation of turn-taking events, or the explicit modeling of prosodic features. Through manipulation of the speech signal, we investigate how these models implicitly utilize prosodic information. We show that these systems learn to utilize various prosodic aspects of speech both on aggregate quantitative metrics of long-form conversations and on single utterances specifically designed to depend on prosody.
\end{abstract}

\section{Introduction}

% Turn-taking
% syntactic/prosodic information
% general, feature agnostic, voice-activity modeling, 
%research questions.

% Copied from VAP
Turn-taking is the fundamental ability of humans to organize spoken interaction, i.e., to coordinate who the current speaker is, in order to avoid the need for interlocutors to listen and speak at the same time~\cite{sacks:74}. A dialog can be viewed as a sequence of turns, constructed through the joint activity of turn-taking between the two speakers. A turn refers to segments of activity where a single speaker controls the direction of the dialog. 

In conversational systems, turn-taking has traditionally been modeled using threshold policies which recognize silences longer than a chosen duration as transition-relevant places. Although these types of models are commonly used, it is well known that they are insufficient for modeling human-like turn-taking \cite{Skantze2021}. Studies of human-human conversation have shown that turns are frequently shifted with a gap of just 200ms \cite{timing:15}, or even with a slight overlap. Thus, given that humans also need some time to prepare a response, it would be infeasible for humans to just use silence as a cue to turn-taking. Instead, it has been suggested that they are able to project turn completions already while the other person is speaking \cite{sacks:74,timing:15,content:15}. In addition, humans produce so-called \textit{backchannels} (short feedback tokens such as "mhm") in a timely manner, often in overlap with the other speaker \cite{Yngve}. 

% Add more sources of what is important for turn-taking (Text vs Prosody)

A common research question in phonetics, psycho-linguistics, and conversational analysis concerns the various cues (including speech, gaze, and gestures) that humans use to detect or project turn-shifts \cite{duncan100829}. When it comes to speech, a common distinction is made between the prosodic (non-lexical) and lexical (textual, syntactic, semantic) components of the speech signal. For example, \citet{ruiter06} argued, based on listening experiments, for the importance of syntactic information over intonation (pitch) in turn-taking, while \citet{phrases} showed that intonation is important when syntactic completion is ambiguous.  However, such studies often require human listening experiments which are costly, anecdotal, and constrained in time resolution and are therefore limited to small amounts of conversational contexts. An alternative approach is to use computational models~\cite{scalable_pitch} to investigate what type of information they are sensitive to. 

\citet{vap} recently proposed Voice Activity Projection, \textbf{VAP}, which is a general, self-supervised turn-taking model. The model incrementally projects the future speech activity of the two speakers directly from raw audio waveforms. The model can be trained on lots of data, without human annotations, and is agnostic with respect to different types of speech information, as it does not depend on explicitly extracted features. This makes the VAP model potentially suitable as a data-driven approach for investigating the role of prosody in turn-taking.

In this work, we train VAP models on a large dataset~\cite{swb, fisher} of dyadic spoken interactions and evaluate it on specific turn-taking metrics, while perturbing the input audio to omit certain sources of prosodic information. We analyze the performance over different tasks to investigate three research questions:

\begin{enumerate}
    \item Do Voice Activity Projection models trained on raw waveforms learn to pick up prosodic information that is relevant to turn-taking?
    \item When/how is prosodic information important for turn-taking predictions?
    \item What is a suitable time resolution for such models to best represent prosody?
    %\item Do data driven turn-taking models converge towards psycho-linguistic theories about the type of information relevant to turn-taking?
\end{enumerate}

\section{Background}
\label{sec:background}

Prosody refers to the non-verbal aspects of speech, including \textit{intonation} (F0/pitch contour), \textit{intensity} (energy), and \textit{duration} (of phones and silences). It has been found to serve many important functions in conversation, including prominence, syntactic disambiguation, attitudinal reactions, uncertainty, topic shifts, and turn-taking \citep{Ward2019}. Studies on both English and Japanese have found that level intonation (in the middle of the speaker's fundamental frequency range) tends to serve as a turn-holding cue, whereas either rising or falling pitch can be found in turn-yielding contexts \cite{gravano101679,local100968,Koiso1998}. When it comes to intensity, studies have found that speakers tend to lower their voices when approaching potential turn boundaries, whereas turn-internal pauses have a higher intensity \cite{gravano101679, Koiso1998}. Regarding duration and speaking rate, \citet{duncan100829} found a ``drawl on the final syllable or on the stressed syllable of a terminal clause" to be a turn-yielding cue (in English). This is also in line with the findings of \citet{local100968}. 

When it comes to lexical information, a very strong cue to turn-taking is of course whether the utterance is syntactically or pragmatically complete \cite{ford_thompson}. Thus, even if prosodic cues can be found near the end of a turn-shift, it is not clear to what extent such cues provide additional information compared to lexical cues, or if they are redundant. In an experiment by \citet{ruiter06}, subjects were asked to listen to a conversation and press a button when they anticipated a turn ending. The speech signal was manipulated to either flatten the intonational contour, or to remove lexical information by low-pass filtering. The results showed that the absence of intonational information did not reduce the subjects' prediction performance significantly, but that their performance deteriorated significantly in the absence of lexical information. From this, they concluded that lexical information is crucial for end-of-turn prediction, but that intonational information is neither necessary nor sufficient. \citet{turngpt} also found that it is possible to build fairly reliable turn-taking models using only lexical information. 

However, it has also been argued that while lexical information is important for turn-taking, there are many cases where a phrase may be syntactically complete, but it is unclear whether the turn is in fact yielded or not \cite{ford_thompson}. To investigate this, \citet{phrases} performed a similar experiment as \citet{ruiter06}, but selected the stimuli so that they contained several syntactic completion points (e.g. ``Are you a student / at this university?"), and where the intonation phrase boundary provided additional cues to whether the turn was yielded or not. They found that subjects indeed made better predictions with the help of intonation and duration. 

Most previous attempts at modeling prosody in turn-taking have been limited in that they (I) only use instances of mutual silence for predicting turn shifts (and therefore do not model projection of turn completion), and (II) only use fairly superficial, hand-crafted features, such as the extracted pitch slope or pitch level right before the pause (e.g., \citealt{gravano101679,meena}). Apart from the problem that such features might be too simplistic, they also typically require speaker normalization of the pitch \cite{zhang-2018-comparison}.  

In this work, we investigate various forms of turn-taking events (including projection of both turn shifts and backchannels). We also use a more agnostic modeling approach, using latent speech representations that are learned in a self-supervised manner and extracted from the raw waveform \cite{cpc}. If our model is indeed able to pick up relevant prosodic information from these representations, it means that we do not have to do any special prosodic feature engineering or speaker normalization.

\section{Voice Activity Projection Model}
\citet{vap} proposed a generic turn-taking model that does not predict specific turn-taking events at specific moments in time. Instead, the model is given the task of Voice Activity Projection (\textbf{VAP}), which means that it has to incrementally predict the future voice activity (\textbf{VA}) of each interlocutor in a dialog. The prediction target at each incremental step is defined by a window of 2 seconds containing the future VA for both speakers. The window is discretized into 8 separate bins (4 for each speaker) where each bin is assigned a value of one if more than half of its frames are active, to produce an 8 bit binary digit, corresponding to 256 unique classes. 

\begin{figure}[ht]
\centering
\includegraphics[width=\linewidth]{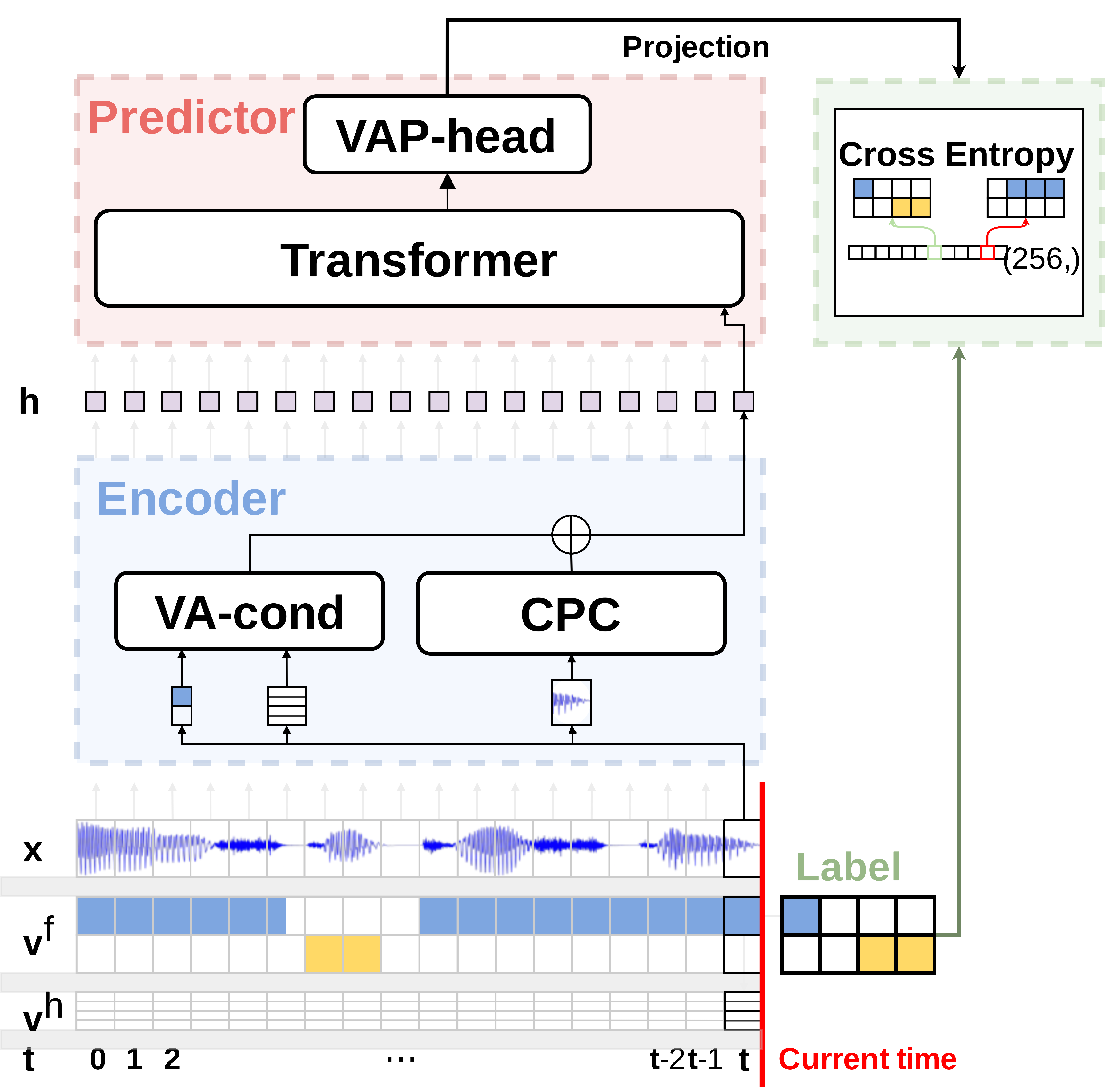}
\caption{The VAP model processes the input features at time $t$. The input to the model is the combined speech waveforms of the two speakers ($x_t$), the VA frames of the window ($v^{f}_t$), and the longer VA history ($v^{h}_t$). The waveform and VA features are processed separately, projected to a common feature space, and added together to produce the predictor input, $h_t$. The predictor consists of a causal transformer feeding into the VAP-head to produce the output projection. The green box illustrates the various outputs of the different models that we compare. Source:~\cite{vap}}
\label{fig:model}
\end{figure}

The VAP model consists of an encoder that processes raw audio waveforms, along with the current VA information, to produce latent representations of a defined frame frequency $f_{enc}$Hz which are then fed into the predictor network. The predictor is a causal sequence network that processes the context available up until the current frame and outputs a probability distribution over the 256 VA classes, see Figure~\ref{fig:model}.

The encoder consists of two sub-modules, a speech module which processes raw waveforms, $x$, specifically a CPC \cite{cpc} model that outputs frame-level representations $h_{speech,t}\in \mathbb{R}^{256}$, at $f_{enc}$ Hz. A second VA module, matching the frame rate of the speech encoder, processes the current VA frame vector $v^f_t\in \{0, 1\}^2$, along with a concise representation of the VA history. The VA history features provide long-ranging contextual information outside of the receptive field of the acoustic model. This history is defined as the activity ratio of speaker A over speaker B for regions of size \{-inf:60, 60:30, 30:10, 10:5, 5:0\} seconds into the past, where 0 is the current time step, resulting in a vector $v^h_t\in\mathbb{R}^5$ with values between 0 and 1, for each frame. The VA module projects the VA features to vectors $h_{va,t},h_{his,t}\in\mathbb{R}^{256}$ which are added to the speech representation $h_{speech,t}$ to produce the encoder output $h_t$, for each frame $t$. The dialog input waveforms are volume normalized, resampled to 16kHz, mixed to a single channel and split into 10s segments (using a 1s overlap). 

The predictor consists of a causal, decoder only, transformer~\cite{transformer}, with linear attention~\cite{alibi}, using a hidden size of 256, 4 layers, 8 heads, and 0.1 dropout. The output of the transformer model is fed to the VAP head, a final linear layer, which outputs logits associated with the 256 VA classes. Since transformer models are powerful but come with the cost that they scale quadratically in compute, with respect to input length, we are interested in whether using a slower frame rate of the sequence model has any significant impact on the turn-taking performance. Following previous work, we utilize a pre-trained CPC~\cite{cpc_across} encoder which produces output representations at 100Hz, and for two of our three models, we include a single additional convolutional layer which projects the representations to 50 and 20Hz. In other words, we train three models which use different frame rates of the predictor.

\subsection{Turn-taking Metrics}
\label{sec:metrics}

\looseness=-1
The Voice Activity Projection in itself is just a distribution of 256 possible futures. However, \citet{vap} also showed how this distribution can be used to predict various turn-taking events as zero-shot classification tasks. We utilize three of these metrics, namely \textit{Shift/Hold}, \textit{Shift-prediction}, and \textit{Backchannel-prediction}, and will briefly explain them here.
%These classification tasks were \textit{Shift vs Hold}, \textit{Shift-prediction}, \textit{Backchannel-prediction} and \textit{Short vs Long}. We will briefly explain them here.

\textbf{Shift/Hold}: 
This metric evaluates how well the model predicts the next speaker during mutual silence, i.e., whether the current speaker will Hold the turn, or whether the turn will Shift to the other speaker. The frames used for evaluation start 50ms into the silence, covering a total of 100ms consecutive frames. 

%The metric depends on two parameters, limiting the analysis towards more ``proper" turn shifts, the \textit{pre-offset} (1s) and \textit{post-onset} (1s), which define areas where only a single speaker can be active, as illustrated in Figure~\ref{fig:events} (Left). 

%\begin{figure}[h]
%\centering
%\includegraphics[width=0.48\linewidth]{figs/event_sh.png}
%\includegraphics[width=0.48\linewidth]{figs/event_bc.png}
%\caption{Turn-taking events}
%\caption{\textbf{Left}: Only a single speaker can be active in regions (green) before/after a mutual silence (red). Shift/Hold are determined by the previous and next speaker (the example illustrates a Shift). \textbf{Right}: The ``backchanneler" cannot be active in the green regions before/after the BC segment.  Source:~\cite{vap}
%}
%\label{fig:events}
%\end{figure}

\textbf{Shift prediction}:
This metric evaluates how well the model can continuously predict an upcoming Shift in the near future, while a speaker is still active. We follow prior work and consider a range of 500ms that covers the end of a VA segment, before a Shift-event (as defined above), as positive samples. Similarly, we sample negative ranges, of the same duration, from regions where a single speaker is active but far away (2s) from any future activity of the other speaker.

\textbf{Backchannel (BC) prediction}:  This metric evaluates how well the model can continuously predict an upcoming BC in the near future (similar to \cite{bc_15, bc_17}). BCs are defined as short and isolated VA segments (\(\leq{1s}\)), preceded by \(\geq{1s}\) of silence and followed by \(\geq{2s}\) of silence by the same speaker. We consider regions of 500ms before a BC as positive samples and the negatives are sampled similarly to the \textit{Shift prediction} metric, with the addition of allowing for non-active segments, i.e., backchannels can be predicted during silences as well. 
%BCs are defined as short and isolated VA segments, as illustrated in Figure~\ref{fig:events} (Right). The metric depends on three parameters: the \textit{bc-duration} (1s), \textit{pre-silence} (1s) and \textit{post-silence} (2s), which controls the degree of isolation and the maximum duration of a segment to be considered a BC. 

\section{Training and Data}

We train three different VAP models with different frame-level frequencies: 20, 50, and 100Hz. We use the combination of two dyadic conversational datasets, Switchboard~\cite{swb} and Fisher\footnote{Because of limited access we only use \href{https://catalog.ldc.upenn.edu/LDC2004S13}{Part 1} of the full corpus.}~\cite{fisher}, resulting in 8288 unique dialogs. We set aside a test set of 5\% (of each dataset) and split the remaining dialogs into a 90/10 train/validation split used for training. We use the AdamW~\cite{adam, opt_decoupled} optimizer and an early stopping criteria on the validation loss with a patience of 10 epochs. The code is implemented in Python using the PyTorch~\cite{pytorch}, PyTorch-Lightning \cite{pl} and Wandb~\cite{wandb} libraries, and are publicly available\footnote{\url{https://github.com/ErikEkstedt/conv_ssl}}.

%\section{Experiments}
%We evaluate the models on the aforementioned turn-taking metrics and perform additional analysis on utterance-lever examples specifically designed to contain ambiguous syntactic information~\cite{phrases} making them suitable for investigating the effect of prosody. 

\begin{figure*}[t]
\centering
\begin{subfigure}[b]{0.31\linewidth}
\centering
\includegraphics[width=\linewidth]{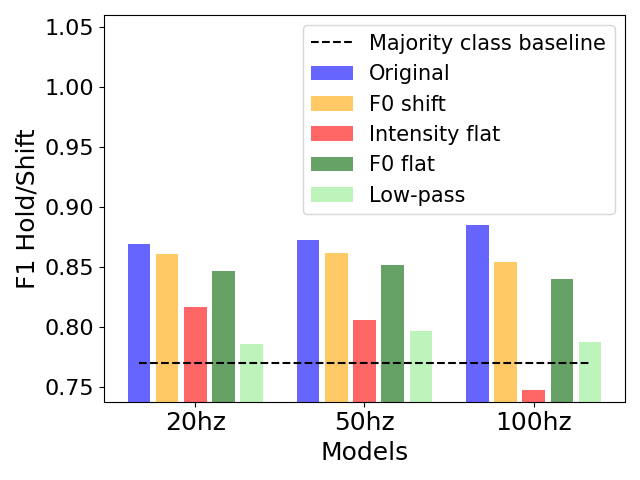}
\caption{Shift/Hold}
\label{fig:agg_hs}
\end{subfigure}
\hfill
\begin{subfigure}[b]{0.31\linewidth}
\centering
\includegraphics[width=\linewidth]{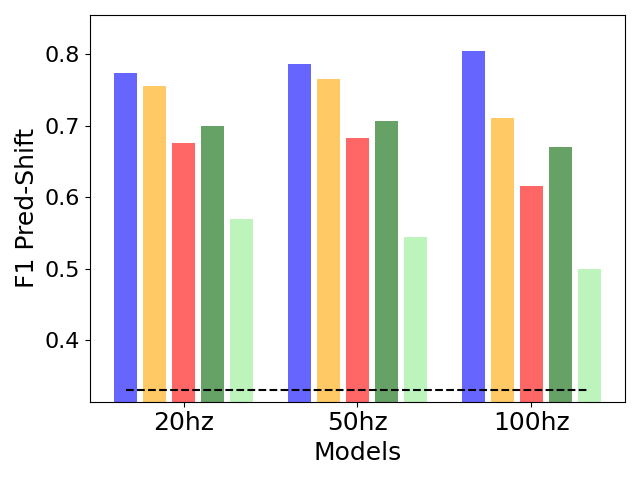}
\caption{Shift prediction}
\label{fig:agg_ps}
\end{subfigure}
\hfill
\begin{subfigure}[b]{0.31\linewidth}
\centering
\includegraphics[width=\linewidth]{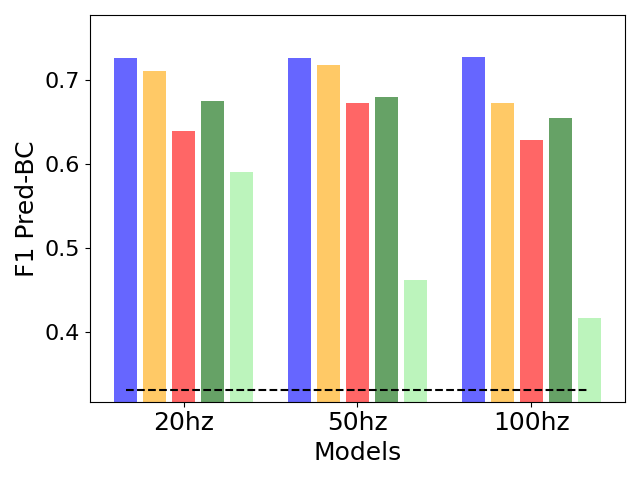}
\caption{Backchannel prediction}
\label{fig:agg_ls}
\end{subfigure}
\caption{Aggregate results for the three tasks on the Switchboard and Fisher test set, depending on model frequency and perturbation. Majority class baseline is shown with the dashed black line.}
\label{fig:agg_results}
\end{figure*}

%\subsection{Data augmentation}
\subsection{Data perturbation}

%In order to investigate the role of prosody in the model's turn-taking predictions, we augment the input audio waveform of the test data in five ways to omit parts of the signal encoding for various prosodic features:
In order to investigate the role of prosody in the model's turn-taking predictions, we perturb the input audio waveform of the test data in five ways to omit parts of the signal encoding for various prosodic features:

\textbf{F0 flat}: the intonation contour is flattened to the average F0 of each speaker and segment.

\textbf{Low pass}: the signal is low pass filtered by down-/up-sampling of the waveform similar to \citet{prosody_rep}. This effectively removes all high-frequency phonetic information, while only the F0 and intensity contours are relatively intact. We use a cutoff frequency of 400Hz across all samples.

\textbf{Intensity flat}: The intensity contour is flattened to the average value of each speaker over all speech frames (as determined by the VA features). We note that this transformation is difficult to perform without including acoustic artifacts despite having access to speech boundaries given by the VA features. Breaths become very loud and the gain inside smaller segments of silence is prominent.

\textbf{Duration average}: Each phone in a segment is scaled to the average duration, of that specific phone, across the dataset. 

\textbf{F0 shift}: The intonation contour is shifted by 90\% of the original value for each speaker over each active speech segment. This should (in theory) not affect the turn-taking predictions. However, we include this perturbation to verify that the transform in itself does not have a too strong effect (e.g., through artifacts).

All perturbations were done using Praat~\cite{praat, parselmouth} and the Torchaudio\footnote{\url{https://pytorch.org/audio}} library.

\section{Aggregate Turn-taking Evaluation}

In this experiment, we evaluate the models on the turn-taking metrics described in Section~\ref{sec:metrics}, on a withheld test set, using the original audio and the respective augmentations, with the exception of \textit{Duration average}\footnote{We do not have access to phone aligned annotations of the datasets.}, listed above. The performance across models and metrics is visualized in Figure~\ref{fig:agg_results}.  We note that the Shift/Hold metric is highly imbalanced, containing a substantially larger amount of holds, indicated by the high baseline weighted F1 ($\approx0.77$). The remaining metrics are balanced by design, resulting in a lower baseline value ($\approx.33$).

% QUESTION:
% When can explain a drop in performance for some augmentation with "the model has learned to use relevant prosodic cues shown by the drop in performance"
% and when e.g. "the 100hz seem too sensitive to artifacts, shown by the drop in perfomance"
% or "the model is really good at turn-taking that despite augmentation it performs well on the task indicating that it has learned to utilize all valid features and when some are omitted it just uses other ones" (well no model have this results but anyway....

The least intrusive augmentation across all models and metrics is, as expected, the \textit{F0 shift} transformation. However, the artifacts introduced still seem to have some effect on the models. Interestingly, it has the greatest impact on the 100Hz model, indicating that a higher frame rate of the predictor model could make it more sensitive to detailed phonetic information disregarded by the slower versions.
 
On the Shift/Hold metric, all models are similarly and substantially impacted by the \textit{Low pass} augmentation, lowering the performance towards baseline performance. This augmentation omits almost all information other than the F0 and intensity contours and shows that the model does rely on more complex cues to predict the next speaker. \textit{F0 flat} interestingly has the least negative effect, across all models (disregarding \textit{F0 shift}). This is surprising, given that pitch seems to be the most frequently used prosodic cue in computational turn-taking models. However, while \textit{Intensity flat} severely affects the 100Hz model, making it worse than the baseline, it has a lesser effect than \textit{Low pass} for the other two.

On the \textit{Shift prediction} and \textit{Backchannel prediction} tasks, where the evaluation point occurs inside of an ongoing utterance, all models are substantially affected by the \textit{Low pass} transform, and the higher the frame rate of the model, the larger the impact. The transformation removes faster phonetic information obfuscating phones, words, and their durations (or boundaries), which are more discernible to models operating on higher frame rates, making the impact variation across models less surprising. However, this variation is greater on the \textit{Backchannel prediction} task, with a large difference of effect between the 20 and 100hz models. The second most impactful perturbation is \textit{Intensity flat}, which indicates, in accordance with the turn-taking literature in general, that shifts and backchannels are preceded by changes (arguably drops) in the intensity contour of the current speaker. 

%We note that flattening the F0 does impact the predictive performance as well but to a slightly lesser degree than the other augmentations. Overall, we note that prosody becomes more important when incrementally predicting upcoming turn-taking events than on the more common reactive approach (Shift vs Hold).

% metric
% [augmentations] model
% [diff from 'Original']
%%%%%%%%%%%%%%%%%%%%%%%%%%%
%%%% f1_hold_shift
% ['low-pass', 'inten', 'flatf0', 'shift', 'regular'] % 20hz
% [0.083, 0.052, 0.022, 0.009, 0.0]
% ['low-pass', 'inten', 'flatf0', 'shift', 'regular'] % 50hz
% [0.076, 0.067, 0.021, 0.011, 0.0]
% ['inten', 'low-pass', 'flatf0', 'shift', 'regular'] % 100hz
% [0.138, 0.098, 0.045, 0.031, 0.0]
%%%% f1_predict_shift
% ['low-pass', 'inten', 'flatf0', 'shift', 'regular'] % 20hz
% [0.203, 0.097, 0.074, 0.018, 0.0]
% ['low-pass', 'inten', 'flatf0', 'shift', 'regular'] % 50hz
% [0.241, 0.103, 0.079, 0.021, 0.0]
% ['low-pass', 'inten', 'flatf0', 'shift', 'regular'] % 100hz
% [0.306, 0.189, 0.134, 0.095, 0.0]
%%%% f1_bc_prediction
% ['low-pass', 'inten', 'flatf0', 'shift', 'regular'] 20hz
% [0.136, 0.086, 0.051, 0.015, 0.0]
% ['low-pass', 'inten', 'flatf0', 'shift', 'regular'] % 50hz
% [0.265, 0.054, 0.046, 0.008, 0.0]
% ['low-pass', 'inten', 'flatf0', 'shift', 'regular'] % 100hz
% [0.312, 0.099, 0.073, 0.055, 0.0]

\section{Utterance-level Analysis}
\label{sec:utt_level}

While the analysis above gives an overall estimate of how important prosody is, it has been hypothesized that prosody is especially important when the semantic/pragmatic completion is ambiguous, as discussed in Section~\ref{sec:background}. To focus their analysis on such situations, \citet{phrases} constructed question templates where a short and a long version, sharing initial lexical information, were recorded through scripted interviews (in Dutch). As an example, a short/long question pair "did you drive here?'' and "did you drive here this morning?'' contain the same initial words up to a common completion point (after the word "here''), which we will refer to as  the \textit{short completion point}, SCP. Note that in order for the listener (or the model) to predict a turn-shift towards the end of the short utterance, but not at the corresponding place in the long utterance, it has to rely on prosody. Through listening experiments, where the participants are asked to press a button when they expect a turn shift, \citet{phrases} found that the reaction time was indeed much faster after the short version than after a long version cut after the SCP. 

\begin{figure}[t]
\centering
\includegraphics[width=\linewidth]{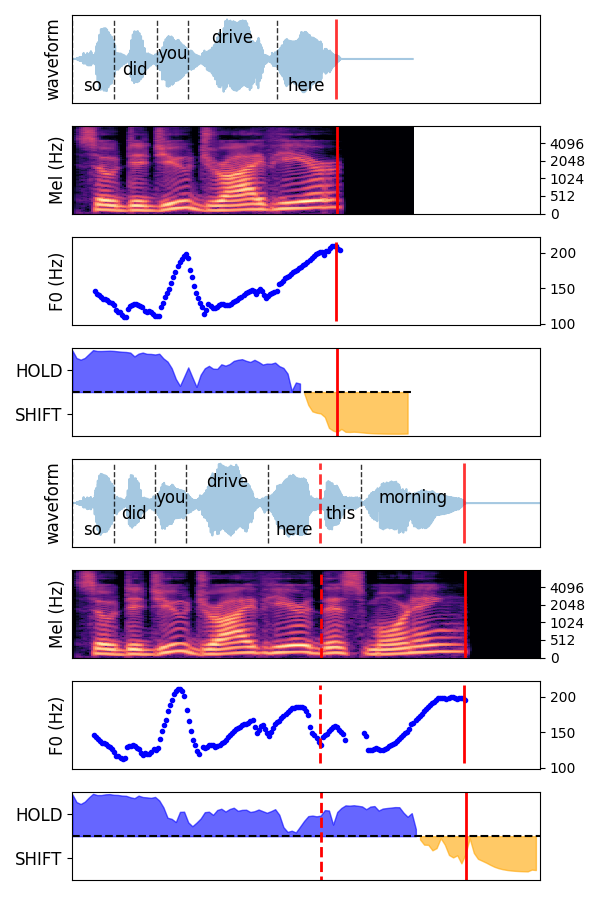}
\caption{A short/long phrase pair. The plots show the waveforms, mel-spectrograms, F0 contours, and the model assigned Shift/Hold comparison, for the short and long versions respectively. The blue color in the bottom plots indicates a majority probability (over 50\%) for Hold whereas the yellow indicates Shift. The short completion point (SCP) is shown as a red dashed line for the long utterance and the filled red line shows the end time of the last word in each utterance.}
%\includegraphics[width=\linewidth]{figs/align_student_female_3.png}
%\caption{A short/long phrase pair. The plots show the waveforms, mel-spectrograms, F0 contours and the model assigned Shift/Hold comparison, for the short and long version respectively. The blue color in the bottom plots indicate a majority probability (over 50\%) for Hold whereas the yellow indicate Shift. The short completion point is located on the word "student'' and the SCP is shown as a red dashed line for the long utterance and the filled red line shows the end time of the last word in each utterance.}
\label{fig:sample_align}
\end{figure}

For our experiments, we created a similar set of 9 long/short utterance pairs in English (see Table \ref{app:tab_phrases} in the Appendix) using the Google TTS\footnote{\url{https://cloud.google.com/text-to-speech}} service and produced 10 versions of each long/short pair using 5 male and 5 female voices. An example of such a pair is visualized in Figure~\ref{fig:sample_align}. In the figure, we have also visualized the VAP model's \textit{Shift prediction}, as described in Section~\ref{sec:metrics}. 

\begin{figure}[t]
\centering
\begin{subfigure}[b]{0.49\linewidth}
\centering
\includegraphics[width=\linewidth]{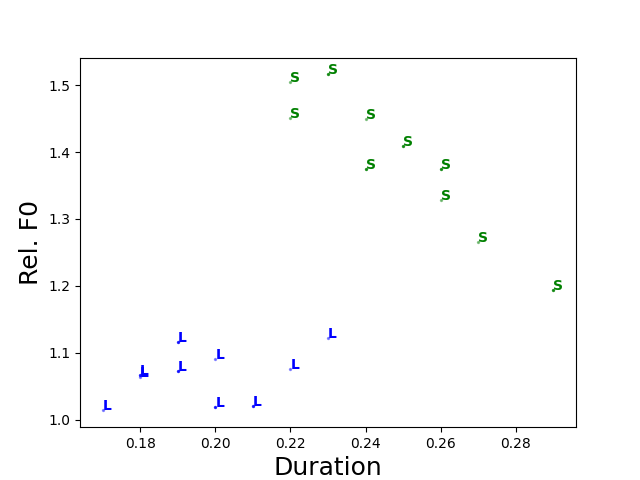}
\caption{Phrase 2. "Psychology''.}
\label{fig:dur_f0_good}
\end{subfigure}
\begin{subfigure}[b]{0.49\linewidth}
\centering
\includegraphics[width=\linewidth]{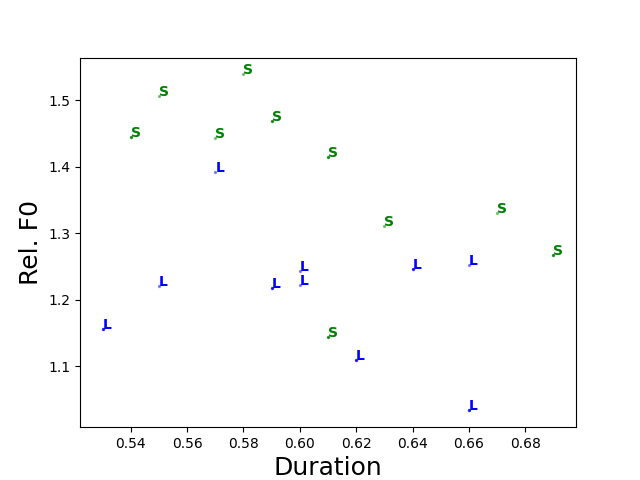}
\caption{Phrase 6. "Live''.}
\label{fig:dur_f0_bad}
\end{subfigure}
\\
\begin{subfigure}[b]{\linewidth}
\centering
\includegraphics[width=\linewidth]{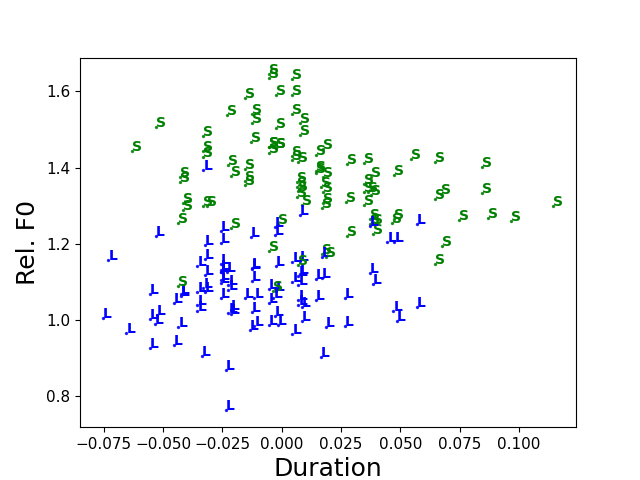}
\caption{All phrases.}
\label{fig:dur_f0_all}
\end{subfigure}
\caption{Duration and maximum relative F0 over the last syllable at the ``short completion point" for the (L)ong and (S)hort versions of the synthesized voices. The x- and y-axis corresponds to mean-shifted duration and relative F0 peak.}
\end{figure}

As can be seen in the figure, for this example, the model correctly assigns a high probability to Hold until towards the end of each utterance, where it changes to Shift. This clearly illustrates the model's ability to project turn shifts before the utterance is complete, and before the large rise in final pitch has actually happened. In addition, we see how the model makes a clear distinction between the two utterances at the short completion point (SCP), where it predicts a Hold for the longer variant. This illustrates that the model is indeed sensitive to prosody, as that is the only information that is different up until that 
%point\footnote{Additional samples are provided at \url{https://erikekstedt.github.io/conv_ssl/}}.
point. Additional samples and visualizations are publicly available\footnote{\url{https://erikekstedt.github.io/conv_ssl/}}.

\begin{figure*}[t]
\centering
\begin{subfigure}[b]{0.31\linewidth}
\centering
\includegraphics[width=\linewidth]{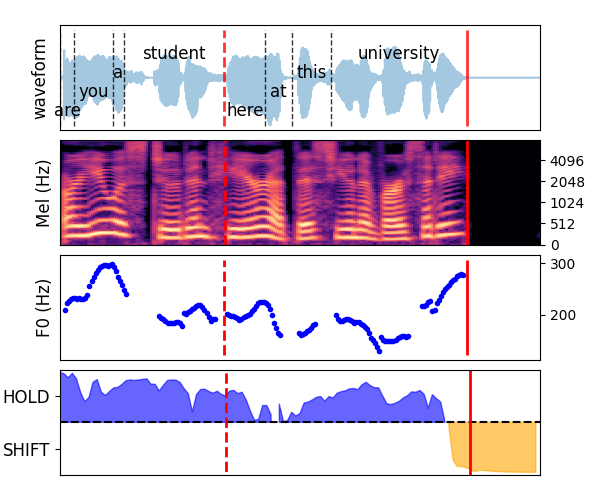}
\caption{Original.}
\label{fig:phrase_orig}
\end{subfigure}
\hfill
\begin{subfigure}[b]{0.31\linewidth}
\centering
\includegraphics[width=\linewidth]{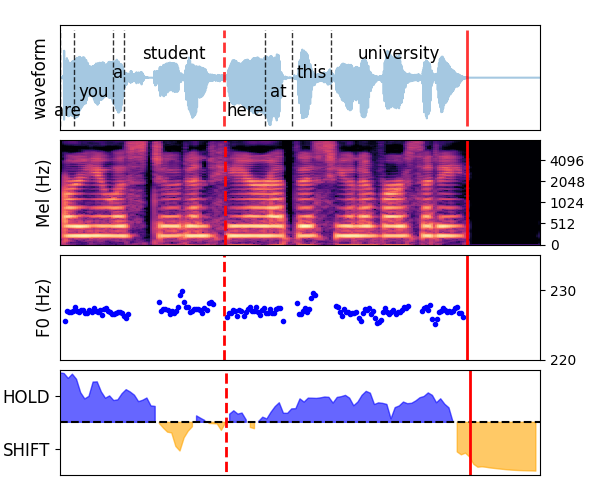}
\caption{F0 flat.}
\label{fig:phrase_f0_flat}
\end{subfigure}
\hfill
\begin{subfigure}[b]{0.31\linewidth}
\centering
\includegraphics[width=\linewidth]{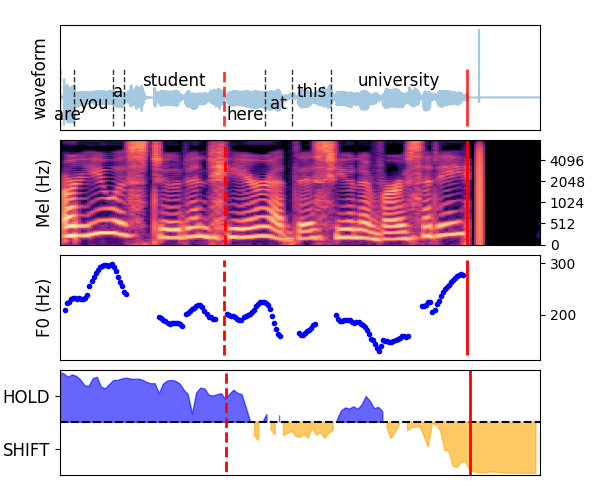}
\caption{Intensity flat}
\label{fig:phrase_in_flat}
\end{subfigure}
\\
\begin{subfigure}[b]{0.31\linewidth}
\centering
\includegraphics[width=\linewidth]{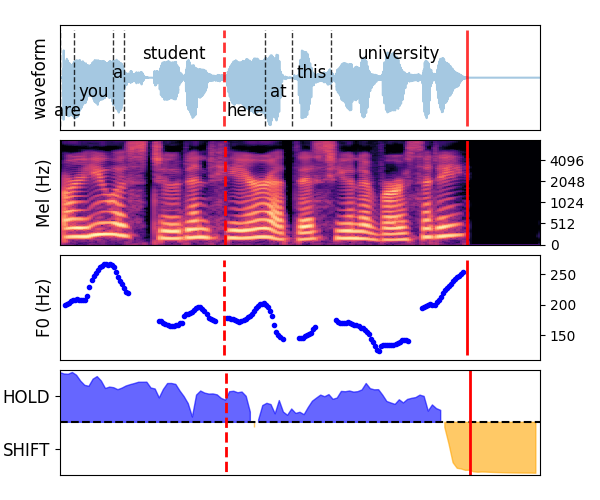}
\caption{F0 shift.}
\label{fig:phrase_f0_shift}
\end{subfigure}
\hfill
\begin{subfigure}[b]{0.31\linewidth}
\centering
\includegraphics[width=\linewidth]{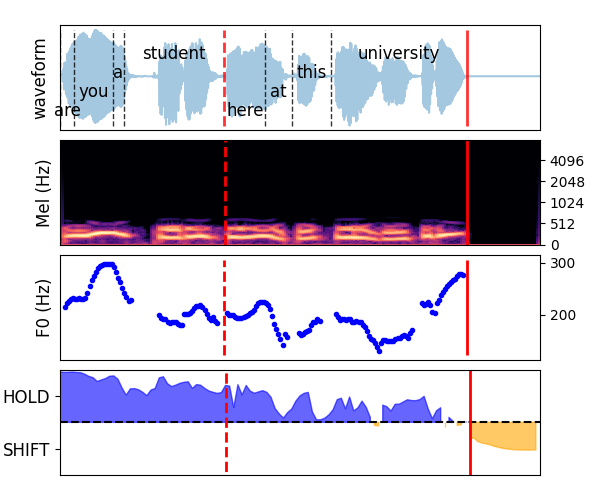}
\caption{Low-pass.}
\label{fig:phrase_low_pass}
\end{subfigure}
\hfill
\begin{subfigure}[b]{0.31\linewidth}
\centering
\includegraphics[width=\linewidth]{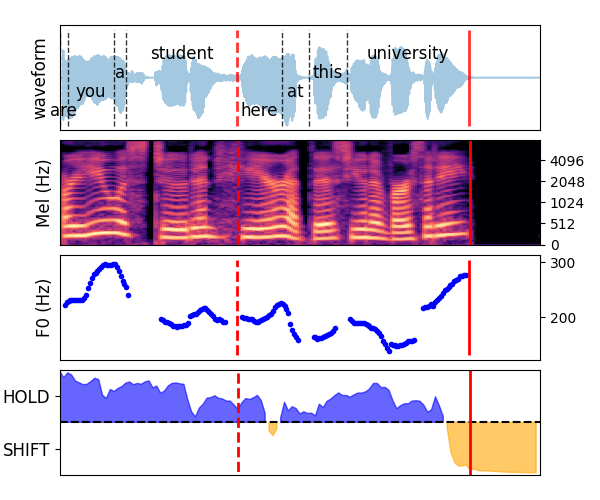}
\caption{Duration average.}
\label{fig:phrase_dur}
\end{subfigure}
\caption{Model output from a female TTS voice saying ``Are you a student here at this university?" (\textbf{long}).}
\label{fig:phrase_long_augmented}
\end{figure*}

Since we rely on artificially generated utterance pairs, we can of course not be certain to what extent they reflect similar prosodic patterns as those generated by humans. We therefore perform a similar analysis of the phrases as \citet{phrases}, by measuring the duration and maximum F0 frequency over the last syllable of the short completion point. In their analysis, they showed that longer duration and a higher rise in F0 are associated with the end of a turn, separating the measures at the SCP of the short phrase from the long, as shown in Figure~\ref{fig:dur_f0_good}. We obtain similar distributions from 4 of our 9 phrases, but note that the others are not as easily separated, but show more uniform distributions for the duration dimension as shown in Figure~\ref{fig:dur_f0_bad}. However, from listening to the phrases, we still consider all recordings natural enough to be included in our further analysis. Although both duration and pitch might sometimes clearly indicate turn-shifts according to the literature, there is no guarantee that this is actually the case for all types of phrases. This indicates that simple models that only track these superficial features might not capture the whole picture. 
%A model with more complex prosodic representations and which can find more complex patterns, on the other hand, might still be able to use prosody to predict turn-taking. 
We provide the mean-shifted duration and relative F0 rise over all generated phrases in Figure~\ref{fig:dur_f0_all}.

We compare the performance of the VAP model on the short and long versions of each phrase to investigate whether it can recognize the prosodic differences and correctly predict the short completion point as either a Hold (long phrase) or a Shift (short phrase). In addition to the original recordings, we include evaluations of the performance on the perturbed versions to investigate whether any specific perturbation changes the predictions of the model more than the others. We use the 50Hz model, as it performs comparably to the 100Hz model on the original audio, while being less affected by the \textit{F0 shift} transform, indicating less sensitivity to arbitrary artifacts introduced by the perturbations. 

The model output on the long version of the phrase ``Are you a student here at this university?", for the various perturbations, is visualized in Figure~\ref{fig:phrase_long_augmented}. Inspection of the original performance in Figure~\ref{fig:phrase_orig} indicates that the model is sensitive to prosodic information and assigns a higher likelihood of a Hold at the SCP located on the word "student''. However, for the \textit{F0 flat} perturbation, in Figure~\ref{fig:phrase_f0_flat}, we note that the model flips and assigns a higher Shift-probability at the SCP, which indicates that if the dynamics of the F0 contour is omitted, the model cannot recognize that the speaker will continue to speak. Interestingly, the \textit{Intensity flat} perturbation also affects the output of the model, but after the SCP is completed. Here, the model does have access to the F0 contour and correctly assigns a larger Hold-probability at the SCP, but then changes prediction to indicate that a Shift is probable following the word "here''. As a final note, the \textit{Low pass} transform, which filters out all phonetic information while keeping both the intensity and F0 contour, does produce predictions close to that of the original audio, while being slightly less certain of a Shift after the entire utterance is completed, as seen in Figure~\ref{fig:phrase_low_pass}. We also provide the corresponding visualizations over the short version of the same speaker and phrase in Figure~\ref{app:fig_phrase_short_augmented} in the Appendix.

% \begin{figure}[t]
% \centering
% \begin{subfigure}[b]{0.48\linewidth}
% \centering
% \includegraphics[width=\linewidth]{figs/phrases/student_female_long_en-US-Wavenet-G_regular.png}
% \caption{Original.}
% \label{fig:phrase_orig}
% \end{subfigure}
% \hfill
% \begin{subfigure}[b]{0.48\linewidth}
% \centering
% \includegraphics[width=\linewidth]{figs/phrases/student_female_long_en-US-Wavenet-G_flat_f0.png}
% \caption{Flat F0.}
% \label{fig:phrase_f0_flat}
% \end{subfigure}
% \\
% \begin{subfigure}[b]{0.48\linewidth}
% \centering
% \includegraphics[width=\linewidth]{figs/phrases/student_female_long_en-US-Wavenet-G_flat_intensity.png}
% \caption{Flat intensity.}
% \label{fig:phrase_in_flat}
% \end{subfigure}
% \hfill
% \begin{subfigure}[b]{0.48\linewidth}
% \centering
% \includegraphics[width=\linewidth]{figs/phrases/student_female_long_en-US-Wavenet-G_only_f0.png}
% \caption{Only F0.}
% \label{fig:phrase_low_pass}
% \end{subfigure}
% \caption{Model output from a female TTS voice saying ``Are you a student at this university?" (\textbf{long}).}
% \label{fig:phrase_long_augmented}
% \end{figure}

\begin{figure}[t]
\centering
\includegraphics[width=\linewidth]{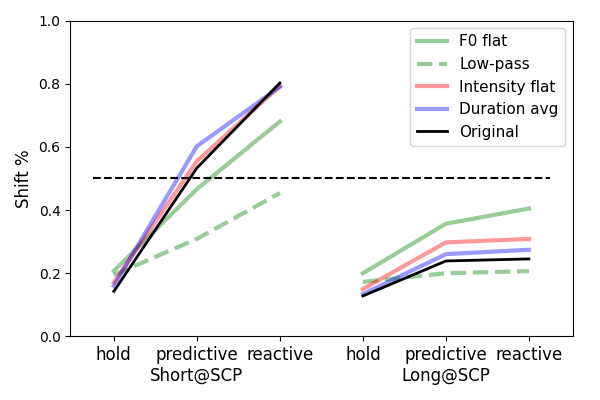}
\caption{Shift probabilities for the 50Hz model on the short completion point over the \textit{hold}, \textit{predictive} and \textit{reactive} regions over all short and long phrases.}
\label{fig:phrase_agg}
\end{figure}

To get an aggregate evaluation of the model across all 9 phrases and 10 voices, we define three regions in each utterance, up until the SCP point (for both long and short phrases), namely \textbf{hold}, \textbf{predictive} and \textbf{reactive}, and measure the average Shift probability predicted by the model in those regions. 
The \textit{hold} region covers the start of the utterances until 200ms before the SCP, where the \textit{predictive} region begins. The final \textit{reactive} region is the very last frame of the SCP where the entire last word (of the short utterance) has been processed. Over the long utterances, the model should consistently predict a low shift probability, given that the speaker will continue their turn, while the shift probabilities should increase over the regions of the short utterances. 
%For the long utterances, the whole region up until the SCP is a \textit{hold} region, where the model should predict a Hold. For the short utterances, the \textit{predictive} region starts 200ms before the SCP, where the model should predict a Shift, whereas the \textit{reactive} region only encompasses the very last frame of the SCP, where the model should also predict a Shift. 
The aggregate model performance over all phrases is visualized in Figure~\ref{fig:phrase_agg}. 

The left part of Figure~\ref{fig:phrase_agg} displays the average Shift probabilities for the points on the SCP for the short phrases (Short@SCP) which preferably should start low and rise consistently. The right part of the figure shows the corresponding performance but on the long phrases (Long@SCP) and should be consistently low, indicating that the speaker will continue their turn. Looking at the non-perturbed signal (Original), and comparing the left and right figures, we see that the model is indeed sensitive to prosody, confirming the anecdotal observation from Figure~\ref{fig:sample_align}. The \textit{Low pass} transform clearly hinders the model from predicting a Shift, indicating that pitch and intensity in themselves are not enough. Among the other perturbations, \textit{F0 flat} seems to have the largest negative effect, which confirms that intonation is important for disambiguating turn completion when lexical information is not enough. Duration seems to be less important, which aligns with the observation in Figure~\ref{fig:dur_f0_all}. 

\section{Conclusion and Discussion}

In this work, we train general computational models of turn-taking, provide analytical methods suitable for evaluating their performance on turn-taking tasks, and investigate how they utilize prosodic information in the speech signal. We investigate the models' reliance on prosody by extending psycho-linguistic experiments designed to measure the effect of prosody on turn-taking in human subjects. We conclude by addressing our three research questions below.

\textit{Do Voice Activity Projection models trained on raw waveforms learn to pick up prosodic information that is relevant to turn-taking?} We apply specific prosodic perturbations to the input signal and show a deterioration across all models on the tasks of turn-taking and backchannel prediction, indicating that prosodic cues are utilized by the models. We note that phonetic information has the largest impact on these measures and that F0 information is less important for turn-taking in general. Even more convincing are perhaps the specific comparisons of the models' ability to predict Shift vs Hold at syntactic completion points, where the lexical information is identical. 
This task requires access to the prosodic dynamics of the signal and should be impossible to distinguish based on lexical information alone. 

\textit{When/how is prosodic information important for turn-taking predictions?} Overall, we show that all models are most sensitive to the \textit{low-pass} transform, indicating that phonetic information is important for turn-taking in general. We note that intensity is at least as important as pitch when applied to actual human long-form conversations, but that pitch plays a more important role for the disambiguation at syntactically equivalent completion points. Interestingly, we note that the importance of duration plays a less important role, indicating that the F0-contour is the most reliable cue in the presence of lexical ambiguity. Another interesting observation in Figure~\ref{fig:phrase_agg} is that even if intonation seems to be the most important individual cue, flattening it does not completely collapse the distinction between turn-holding and turn-yielding. Thus, there must also be redundant information in intensity and/or duration. This shows that prosody is indeed a complex set of signals, which the model has captured. 

\textit{What is a suitable time resolution for such models to best represent prosody?} In our analysis of the turn-taking metrics, we note a negligible performance degradation when decreasing the frame rate of the predictor model. We note that high-frequency models tend to focus more on phonetic information, indicated by their sensitivity to the \textit{Low pass} transformation. The faster models seem more sensitive to general acoustic artifacts, as indicated by the larger performance drop on the \textit{F0 shift} perturbation, which should not have an impact on turn-taking cues in general. Overall, we favor the slower models given their lower memory and computational requirements, their robustness, and comparable performance. 

It should be noted that the models were not trained on perturbed versions of the data, which include highly unnatural speech (i.e., no humans speak with a perfect flat intonation contour). Thus, the evaluations of Section \ref{sec:utt_level} can be considered out-of-distribution. Nevertheless, it is interesting that for many of these perturbations, the models still perform relatively well. Also, the drops in performance are typically in line with what could be expected from the literature. For future work, it could be valuable to train multiple models, on data with different prosodic perturbations, and compare their performance for further analysis.
Another interesting approach could be to identify actual instances of syntactically ambiguous phrases, rather than relying on TTS. Moreover, it would be interesting to include a larger linguistic context, and investigate whether the importance of prosody decreases.
%For future work, it would be valuable to identify actual instances of syntactically ambiguous phrases, rather than relying on TTS. Moreover, it would be interesting to include a larger linguistic context, and investigate whether the importance of prosody decreases.
% add Robustness to perturbations and the fact that we did not train on perturbed data and not that testing on OOD data could lead to degraded results

\section{Acknowledgments}
This work was supported by Riksbankens Jubileumsfond (RJ), through the project \textit{Understanding predictive models of turn-taking in spoken interaction} (P20-0484), as well as the Swedish Research Council, through the project \textit{Prediction and Coordination for Conversational AI} (2020-03812).

\bibliography{anthology,custom}
\bibliographystyle{acl_natbib}

\clearpage

\appendix
\onecolumn
\section{Appendix}
\label{sec:appendix}

\label{sec:appendix_phrases}
\begin{table*}[ht]
\caption{The 9 phrases used in the utterance-level analysis.}
\label{app:tab_phrases}
\centering
\begin{tabular}{p{1cm} | p{0.3\linewidth} | p{0.5\linewidth}}
Item & Short & Long \\
\hline
1 & Are you a student? & Are you a student \textbf{here at this university?} \\
2 & Do you study psychology? & Do you study psychology \textbf{here at this university?} \\
3 & Are you a first-year student? & Are you a first-year student \textbf{here at this university?} \\
4 & So do you play basketball? & So do you play basketball \textbf{on Thursdays?} \\
5 & Have you participated in any experiments before? & Have you participated in any experiments before \textbf{here at this university?} \\
6 & Do you live by yourself? & Do you live by yourself \textbf{or with someone else?} \\
7 & So you work on the side? & So you work on the side \textbf{in a supermarket in addition to your studies?} \\
8 & Did you come here by bike? & Did you come here by bike \textbf{this morning?} \\
9 & Did you drive here? & Did you drive here \textbf{this morning?} \\
\end{tabular}
\end{table*}

\begin{figure*}[ht]
\centering
\begin{subfigure}[b]{0.31\linewidth}
\centering
\includegraphics[width=\linewidth]{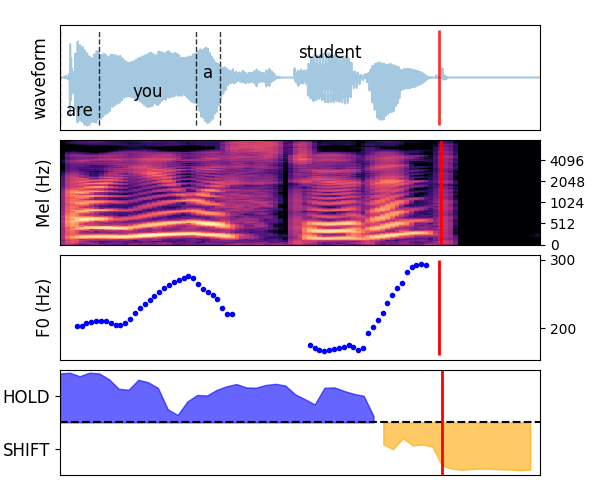}
\caption{Original.}
\label{fig:short_orig}
\end{subfigure}
\hfill
\begin{subfigure}[b]{0.31\linewidth}
\centering
\includegraphics[width=\linewidth]{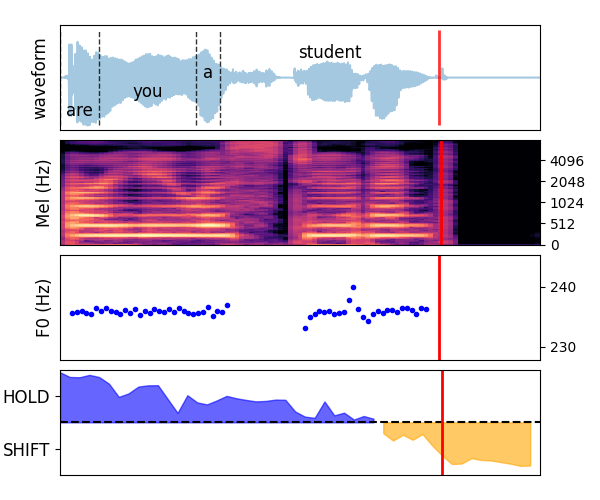}
\caption{Flat F0.}
\label{fig:short_f0_flat}
\end{subfigure}
\hfill
\begin{subfigure}[b]{0.31\linewidth}
\centering
\includegraphics[width=\linewidth]{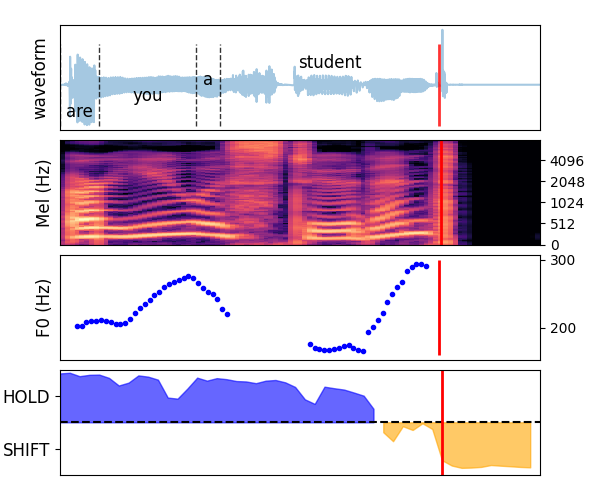}
\caption{Flat intensity.}
\label{fig:short_in_flat}
\end{subfigure}
\\
\begin{subfigure}[b]{0.31\linewidth}
\centering
\includegraphics[width=\linewidth]{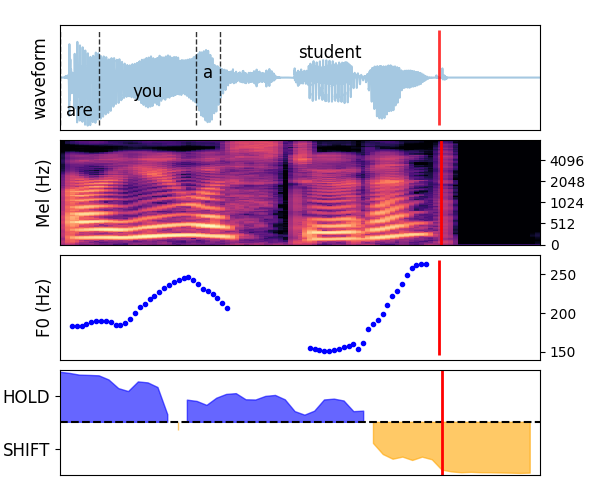}
\caption{F0 shift.}
\end{subfigure}
\hfill
\begin{subfigure}[b]{0.31\linewidth}
\centering
\includegraphics[width=\linewidth]{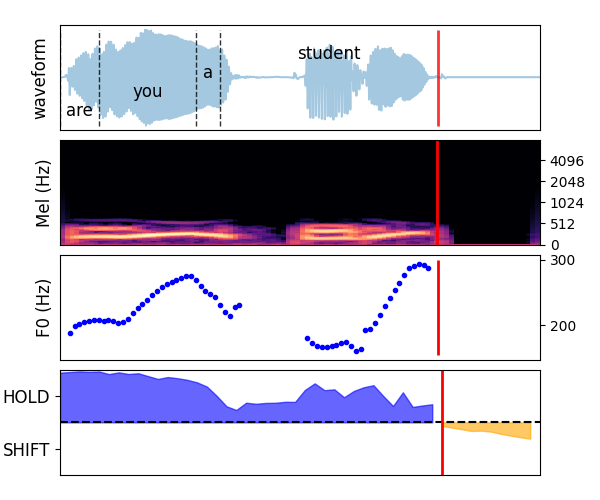}
\caption{Low-pass.}
\label{fig:short_low_pass}
\end{subfigure}
\hfill
\begin{subfigure}[b]{0.31\linewidth}
\centering
\includegraphics[width=\linewidth]{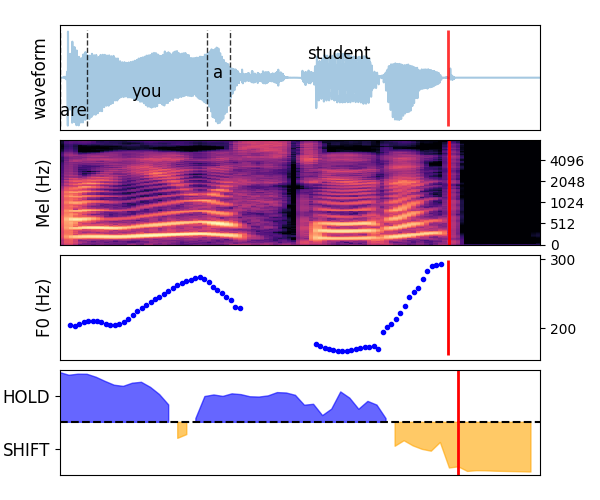}
\caption{Duration avg.}
\end{subfigure}
\caption{Model output from a female TTS voice saying ``Are you a student?" (\textbf{short}).}
\label{app:fig_phrase_short_augmented}
\end{figure*}

\end{document}